\documentclass[conference]{IEEEtran}
\ifCLASSOPTIONcompsoc
  \usepackage[nocompress]{cite}
\else
  \usepackage{cite}
\fi
\ifCLASSINFOpdf
\else
\fi
\usepackage{cite}
\usepackage{amsmath,amssymb,amsfonts}
\usepackage{algorithmic}
\usepackage{graphicx}
\usepackage{textcomp}
\usepackage{xcolor}
\usepackage{fancyhdr}

\def\BibTeX{{\rm B\kern-.05em{\sc i\kern-.025em b}\kern-.08em
    T\kern-.1667em\lower.7ex\hbox{E}\kern-.125emX}}
    
\usepackage{booktabs} 
\usepackage{multirow}
\usepackage{booktabs}
\usepackage{graphicx}  
\usepackage{diagbox}
\usepackage[braket]{qcircuit} 
\usepackage{threeparttable} 
\usepackage{stfloats} 
\usepackage{enumitem} 
\usepackage{fancyheadings} 
\usepackage{wrapfig} 

\usepackage{subcaption,booktabs}

\begin{document}
%


\title{AutoQC: Automated Synthesis of Quantum Circuits Using Neural Network}

\author{
\IEEEauthorblockN{Kentaro Murakami}
\IEEEauthorblockA{Kyushu University\\
Fukuoka, Japan\\
murakami.kentaro.514@gmail.com}
\and
\IEEEauthorblockN{Jianjun Zhao}
\IEEEauthorblockA{Kyushu University\\
Fukuoka, Japan\\
zhao@ait.kyushu-u.ac.jp
}
}


%


\maketitle

\begin{abstract}
While the ability to build quantum computers is improving dramatically, developing quantum algorithms is very limited and relies on human insight and ingenuity. Although a number of quantum programming languages have been developed, it is challenging for software developers who are not familiar with quantum computing to learn and use these languages. It is, therefore, necessary to develop tools to support developing new quantum algorithms and programs automatically.
This paper proposes $AutoQC$, an approach to automatically synthesizing quantum circuits using the neural network from input and output pairs. We consider a quantum circuit a sequence of quantum gates and synthesize a quantum circuit probabilistically by prioritizing with a neural network at each step.
The experimental results highlight the ability of $AutoQC$ to synthesize some essential quantum circuits at a lower cost.
\end{abstract}

\textbf{\itshape Keyword -- Quantum computing; program synthesis; neural network; software development}

%
\IEEEpeerreviewmaketitle

\section{Introduction}\label{sec:introduction}
In recent years, the practical use of quantum computers has been advancing. While the ability to build quantum computers is improving dramatically\cite{arute2019quantum}, the ability to develop quantum algorithms is very limited and relies on human insight and ingenuity. 
Although a number of quantum programming languages such as Q\#~\cite{svore2018q}, Scaffold~\cite{abhari2012scaffold}, OpenQASM~\cite{cross2017open}, ProjectQ~\cite{haner2016high}, and Qiskit~\cite{gadi_aleksandrowicz_2019_2562111} have been developed, it is still challenging for software developers who are not familiar with quantum computing to learn and use these languages. Therefore, it is necessary to develop tools to support the automated development of new quantum algorithms and circuits.

Program synthesis~\cite{gulwani2017program} is the task of automatically constructing executable code fragments, given a user's intent, using various forms of constraints, such as input-output examples, demonstrations, and natural languages. Program synthesis has direct applications~\cite{singh2013automated,gulwani2010dimensions} for various users such as students and teachers, software developers, and algorithm designers in classical computing.
Recently, the idea of program synthesis has also been applied to quantum computing to support the synthesis of quantum circuits during the quantum compiling process~\cite{di2016parallelizing,davis2020towards}. In this thread of research, quantum circuit synthesis is regarded as a process in which an arbitrary unitary operation is decomposed into a sequence of gates from a universal set, typically one which a quantum computer can implement both efficiently and fault-tolerantly. That is, given an arbitrary quantum circuit $C$ and a universal gate set $G$, one seeks to find a decomposition $U_{k}U_{k-1} \ldots U_{2}U_{1} = C$, where $U_{i} \in G$. 
However, there is little research on program synthesis in the quantum computing domain that aims to help software developers synthesize quantum circuits and code pieces. 

This paper proposes $AutoQC$, an approach to automatically synthesizing quantum circuits using the neural network from input and output pairs. In contrast, to support the quantum compilation process, our goal is to help software developers automatically synthesize a quantum circuit or algorithm using machine learning from a problem specification provided as an example of assumed input/output pairs. As the first step, this paper focuses on quantum circuits with classical inputs and outputs, i.e., the problem to be solved by the quantum algorithm is classical bits.

Our paper makes the following contributions:

\begin{itemize}
    \item \textbf{Synthesis Approach.} It presents a novel automated synthesis approach that uses a neural network to explore the space of possible quantum circuit designs by using the input-output pairs.
    \item \textbf{Tool Support}: It implements a synthesis tool called $AutoQC$ to support the automated synthesis of quantum circuits.
    \item \textbf{Experimental Results:} It presents experimental results using $AutoQC$ to synthesize seven quantum circuits from their input-output pairs automatically. The results highlight the ability of $AutoQC$ to synthesize some important quantum circuits.
\end{itemize}

The rest of the paper is organized as follows. Section~\ref{sec:preliminaries} briefly introduces quantum computers and their computations as a prerequisite for explaining our approach. 
Section~\ref{sec:synthesis} describes our approach for automatically synthesizing quantum circuits, and Section~\ref{sec:experimental} describes some experiments on synthesizing quantum circuits using our approach. Section~\ref{sec:related work} discusses related work, and concluding remarks are given in Section~\ref{sec:conclusion}.

\section{Background}\label{sec:preliminaries}

First, we introduce some basic information on quantum computation. More detailed reading material can be found in the book by Nielsen and Chuang~\cite{nielsen2002quantum}.

\subsection{Quantum State and Quantum Bit}

In a classical computer, information is represented by two states, 0 and 1. For example, the two states are represented by the on/off state of a switch, the state in which a charge is accumulated and the state in which it is not, and the high/low voltage. On the other hand, quantum mechanics allows the superposition of two different states, so the "quantum" bit, the smallest unit of information in the quantum world, can be described using two complex numbers $\alpha$ and $\beta$. $\alpha$ and $\beta$ are called complex probability amplitudes and represent the weight with which the 0 and 1 states are superimposed. The reason why $\alpha$ and $\beta$ are complex numbers is that in quantum theory, discrete quantities such as 0 and 1 have the properties of waves and can interfere.

The states corresponding to the 0 and 1 of the classical bits can be represented using Dirac's bracket notation, which is a simplified representation of the column vector, as follows:
$$
  \ket{0} = \left(
  \begin{array}{cc}
  1 \\
  0 \\
  \end{array}
  \right), 
  \ket{1} = \left(
  \begin{array}{cc}
  0 \\
  1 \\
  \end{array}
  \right)
$$

Since $\ket{0}$ and $\ket{1}$ form an orthonormal basis, the qubit state $\ket{\psi}$ can be represented by a linear combination of $\ket{0}$ and $\ket{1}$.
$$
  \ket{\psi} = \alpha \ket{0} + \beta \ket{1} = \left(
  \begin{array}{cc}
       \alpha \\
       \beta
  \end{array}
  \right)\\
$$

\subsubsection{Probability Amplitude}
In quantum mechanics, the observer cannot directly interfere with the complex probability amplitude, and the probability of 0 or 1 is determined only when a measurement operation is performed. The complex probability amplitude influences the probability distribution of the measurement result. The probability that the measurement result will be 0 or 1 is denoted as $p_0$ and $p_1$, respectively. Then these probabilities are expressed as the square of the absolute value of the complex probability amplitude:
$$
    p_0 = |\alpha|^2,  p_1 = |\beta|^2
$$

A normalization condition $|\alpha|^2 + |\beta|^2 = 1$ is imposed to make the sum of the probabilities be equal to 1.

When a measurement is performed, the quantum state transitions to the state corresponding to the measurement result. Specifically, when the measurement result is 0, the state transitions to $\ket{0}$; when it is 1, the state transitions to $\ket{1}$. This measurement is called a projective measurement in the orthonormal basis $\ket{0}$ and $\ket{1}$.

\subsubsection{Multiple Qubits}
When there are n qubits, their states are represented using tensor products.

The states of the n classical bits are represented by n 0 or 1 number, with a total of $\text{2}^n$ patterns. Since quantum mechanics allows the superposition of all these patterns, the state of n qubits can be described by $\text{2}^n$ complex probability amplitudes. Each complex probability amplitude represents which bit sequence is superimposed with which weight.

$$
    \ket{\psi} = c_{00\ldots0}\ket{00\ldots0} + c_{00\ldots1}\ket{00\ldots0} + \cdots + c_{11\ldots1}\ket{11\ldots1} = 
$$
$$  
    \left(
  \begin{array}{cc}
       c_{00\ldots0} \\
       c_{00\ldots1} \\
       \vdots \\
       c_{11\ldots1}
  \end{array}
  \right)
$$

The complex probability amplitudes are assumed to be normalized:
$$
\sum_{i_i, \ldots, i_n}|c_{i_i, \ldots, i_n}|^2 = 1
$$

Then, when we measure the quantum state of these n qubits, the bit sequence $i_i, \ldots, i_n$ is obtained randomly with probability $p_{i_i, \ldots, i_n} = |c_{i_i, \ldots, i_n}|^2$, and the state after the measurement is $\ket{i_i, \ldots, i_n}$.

Thus, the state of n-qubits must be described by a complex vector of $2^n$ dimensions, which is exponentially large with respect to n. This is where the difference between a classical bit and a qubit becomes apparent. The operations on an n-qubits system are then represented as a $2^n \times 2^n$ dimensional unitary matrix.

\subsection{Quantum Gates}\label{chap:quantum gate}

%
We first describe how operations on qubits are represented, which is closely related to the following properties of quantum mechanics:

\begin{itemize}
    \item \textbf{Liniarity:} The time evolution of a quantum state is always linear with respect to the superposition of the states. Since the quantum state of a qubit is represented as a normalized two-dimensional complex vector, operations on a qubit, linear operations, are represented by a $2 \times 2$ complex matrix.
    \item \textbf{Unitarity:} Furthermore, using the normalization condition that the sum of the probabilities is always 1, we can derive the following constraint on quantum operations $U$:
        $$
            U^{\dagger}U = UU^{\dagger} = I
        $$
        In other words, quantum operations are represented by unitary matrices.
\end{itemize}

Here, let us review the terminology. In quantum mechanics, a linear transformation of a state vector is called an operator, and the term operator refers to any linear transformation that is not necessarily unitary. On the other hand, linear transformations that satisfy the unitarity are called quantum gates. A quantum operation can be thought of as an operator on a quantum state that is physically feasible, at least theoretically.

\subsubsection{Single Qubit Gates}
The first single-qubit gate is the X gate, corresponding to NOT in classical computers. The X gate is one of the Pauli operators of basic quantum arithmetic and is a matrix such that

\begin{equation*}
      \text{X} = \left(
      \begin{array}{cc}
          0 & 1 \\
          1 & 0 \\
      \end{array}
      \right)
  \end{equation*}

X gate acts as follows:
$$
    X\ket{0} = \ket{1}, X\ket{1} = \ket{0}
$$

Next is the V and $\text{V}^\dagger$ gates, both of which are a square root of the X gate as described below:

$$
    \text{V} = X^{\frac{1}{2}} = \frac{i+1}{2}\left(
      \begin{array}{cc}
          1 & -i \\
          -i & 1 \\
      \end{array}
      \right)
$$
$$
    \text{V}^\dagger = X^{-\frac{1}{2}} = \frac{i+1}{2}\left(
      \begin{array}{cc}
          1 & i \\
          i & 1 \\
      \end{array}
      \right)
$$

\subsubsection{Multi-Qubit Gate}
Controlled gates are important for operations on two-qubit. Take the Controlled-X gate (CX) as an example, the matrix is written as follows:

\begin{equation*}
      \text{CX} = \left(
      \begin{array}{cccc}
          1 & 0 & 0 & 0 \\
          0 & 1 & 0 & 0 \\
          0 & 0 & 0 & 1 \\
          0 & 0 & 1 & 0
      \end{array}
      \right)
  \end{equation*}
  
CX gate takes two qubits as arguments, a control qubit and a target qubit, and has the following properties:

\begin{itemize}
    \item The CX gate does not modify the control qubit.
    \item If the control qubit is 0, no change is made to the target qubit.
    \item If the control qubit is 1, the X-gate is applied to the target qubit.
\end{itemize}

Such an action corresponds to XOR in classical computation. Therefore, the CX gate can be regarded as a reversible version of XOR.

In the same way, we can consider the controlled gate for the V gate and $\text{V}^\dagger$ gate and are denoted as CV and $\text{CV}^\dagger$.


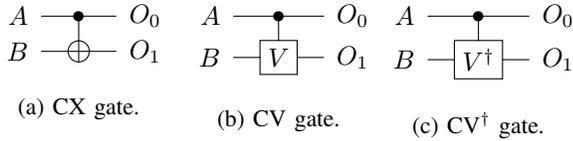
\begin{figure}[bth]
    \begin{tabular}{ccc}
        \begin{minipage}[t]{.3\linewidth}
            \centering
            \[
            \Qcircuit @C=1em @R=.8em{
            \lstick{A} & \ctrl{1} &  \rstick{O_0} \qw \\
            \lstick{B} & \targ    &  \rstick{O_1} \qw
            }\]
            \subcaption{CX gate.}
            \label{fig:cx}
        \end{minipage}%
        \begin{minipage}[t]{.3\linewidth}
            \centering
            \[
            \Qcircuit @C=1em @R=.8em{
            \lstick{A} & \ctrl{1} &  \rstick{O_0} \qw \\
            \lstick{B} & \gate{V}    &  \rstick{O_1} \qw
            }\]
            \subcaption{CV gate.}
            \label{fig:cv}
        \end{minipage}%
        \begin{minipage}[t]{.3\linewidth}
            \centering
            \[
            \Qcircuit @C=1em @R=.8em{
            \lstick{A} & \ctrl{1} &  \rstick{O_0} \qw \\
            \lstick{B} & \gate{V^\dagger}    &  \rstick{O_1} \qw
            }\]
            \subcaption{$\text{CV}^\dagger$ gate.}
            \label{fig:cv+}
        \end{minipage} 
    \end{tabular}
    \caption{Symbols of quantum gates CX, CV, and $\text{CV}^\dagger$.}
    \label{fig:symbols}
\end{figure}

\subsection{Quantum Circuit}

\begin{figure}[bth]
    \centering
    \[
    \Qcircuit @C=1.2em @R=1em{
        \lstick{A} & \ctrl{3} & \qw      & \qw      & \ctrl{2} & \qw      & \qw      & \rstick{O_0} \qw \\
        \lstick{B} & \qw      & \ctrl{2} & \qw      & \qw      & \ctrl{1} & \qw      & \rstick{O_1} \qw \\
        \lstick{C} & \qw      & \qw      & \ctrl{1} & \targ    & \targ    & \ctrl{1} & \rstick{O_2} \qw \\
        \lstick{D} & \gate{V} & \gate{V} & \gate{V} & \qw      & \qw      & \gate{V^\dagger} & \rstick{O_3} \qw \\
    }\]
    \caption{HNG~\cite{HNG}. A, B, C and D are the inputs, and $O_0=A$, $O_1=B$, $O_2=A\oplus B\oplus C$ and $O_3=(A\oplus B)C\oplus AB\oplus D$ are the outputs.}
    \label{fig:HNG}
\end{figure}
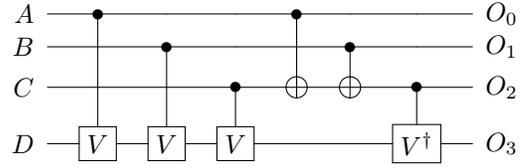

A quantum circuit is a symbolic representation of the structure of the basic operations that realize a given operation, as shown in Figure~\ref{fig:HNG}. The basic operations are described by symbols as shown in Figure~\ref{fig:symbols}, with the operations as described above. Each line running corresponds to a qubit, and the symbols on the lines are the basic operations applied to that qubit, arranged in chronological order from left to right. According to the quantum circuit, the desired output state can be reached by performing the basic operations in order from left to right. Therefore, a quantum circuit can represent a sequence of execution instructions.

\section{Quantum Circuit Synthesis Using Neural Network}\label{sec:synthesis}

A quantum system performs reversible transformations on a quantum state composed of qubits. In principle, a sparse quantum circuit can be generated automatically if a decision procedure exists that can directly reason about the algebraic representation of the quantum circuit and search for a quantum circuit that can perform the given operation. The search for a correct quantum algorithm requires a search for the correct topology of the network representing the quantum circuit and the correct quantum gates to be placed in this topology. However, existing combinatorial search algorithms cannot search high-dimensional spaces with complex values, making this direct approach impractical.
Therefore, we propose to use a neural network to perform an efficient search. In this section, we describe the specific procedure of the proposed approach. The objective of the proposed approach is to generate a quantum circuit that satisfies the conditions from the input-output pairs of the quantum circuit to be generated.

\subsection{Input and Output}
\begin{figure}[h]
\large{
    \centering
    \[
    \Qcircuit @C=1em @R=1em{
        \lstick{A} & \gate{H} & \ctrl{1} &  \rstick{O_0} \qw \\
        \lstick{B} & \qw      & \targ    &  \rstick{O_1} \qw
    }\]
    \caption{A quantum circuit that creates an entanglement state, also called bell state. For example, applying this circuit to $\ket{00}$, there is a 0.5 probability of observing a state of $\ket{00}$ and a 0.5 probability of observing a state of $\ket{11}$. Moreover, there is a 0 probability that observing a state of $\ket{01}$ or $\ket{10}$.}
    \label{fig:bell}
}
\end{figure}
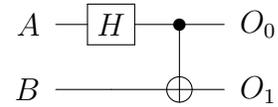

First, the input to the synthesis algorithm is the final state after applying the desired quantum circuit to the $2^n$ computational basis. Then, the quantum gates are generated backward to return the final state to the initial state. The restriction of the quantum circuit specification description to $2^n$ bases of computation is based on the insight that most quantum algorithms are expected to work with a finite number of classical inputs. Taking the quantum circuit in Figure~\ref{fig:bell} as an example, we describe the inputs and outputs of the approach in detail.

The number of qubits in the quantum circuit shown in Figure~\ref{fig:bell} is 2, so the size of the input-output vector is $2^2=4$. The elements of this vector are complex numbers, and the four inputs are the computational basis from 0 to 3, resulting in four complex vectors, as shown below.
\begin{eqnarray*}
    \ket{00} &=& \left(
      \begin{array}{cccc}
          1 & 0 & 0 & 0
      \end{array}
      \right)^T \\
    \ket{01} &=& \left(
      \begin{array}{cccc}
          0 & 1 & 0 & 0
      \end{array}
      \right)^T \\
    \ket{10} &=& \left(
      \begin{array}{cccc}
          0 & 0 & 1 & 0
      \end{array}
      \right)^T \\
    \ket{11} &=& \left(
      \begin{array}{cccc}
          0 & 0 & 0 & 1
      \end{array}
      \right)^T
\end{eqnarray*}

We apply the quantum circuit U shown in Figure~\ref{fig:bell} to the above inputs. Then, we get the so-called bell state, as shown below.

\begin{eqnarray*}
    U\ket{00} &=& \left(
      \begin{array}{cccc}
          \frac{1}{\sqrt{2}} & 0 & 0 & \frac{1}{\sqrt{2}}
      \end{array}
      \right)^T \\
    U\ket{01} &=& \left(
      \begin{array}{cccc}
          0 & \frac{1}{\sqrt{2}} & \frac{1}{\sqrt{2}} & 0
      \end{array}
      \right)^T \\
    U\ket{10} &=& \left(
      \begin{array}{cccc}
          \frac{1}{\sqrt{2}} & 0 & 0 & -\frac{1}{\sqrt{2}}
      \end{array}
      \right)^T \\
    U\ket{11} &=& \left(
      \begin{array}{cccc}
          0 & \frac{1}{\sqrt{2}} & -\frac{1}{\sqrt{2}} & 0
      \end{array}
      \right)^T
\end{eqnarray*}

The input to the algorithm is the state after applying the quantum circuit to the computational basis as described above. Through a search, the synthesis algorithm generates a quantum circuit that satisfies at least the input-output conditions of the computational basis. There are three types of quantum gates that are used by our synthesis approach: CX gate, CV gate, and $\text{CV}^\dagger$ gate. Here, we used the same quantum gates as previous work~\cite{various_gate} proposed quantum versions of classical circuits. The purpose is to compare the cost of quantum circuits designed by humans and synthesized automatically by our synthesis approach.


\subsection{The Synthesis Algorithm}
We next describe the details of our synthesis algorithm. We represent a quantum circuit as a sequence of quantum gates. Then, the following sequence would be the HNG~\cite{HNG} shown in Figure~\ref{fig:HNG}.
$$
\text{CV}(0, 3), \text{CV}(1, 3), \text{CV}(2, 3), \text{CX}(0, 2), \text{CX}(1, 2), \text{CV}^\dagger(2, 3)
$$

In our approach, we generate such a sequence of quantum gates from back to front. At each step, quantum gates are generated stochastically. Here, the word "stochastically" does not mean random but rather some prioritization. The neural network assigns this priority. To demonstrate the effectiveness of this prioritization, we have also conducted an experiment of random search synthesis. In the following, we explain the random search-based approach first.

\subsubsection{Random Search}
In a random search, the quantum gate is chosen completely randomly at each step. This procedure is repeated a certain number of times to generate a quantum circuit. The generated quantum circuit is checked to see if it satisfies the input and output conditions, and if not, another quantum circuit is generated.

The reason for not searching all possible quantum circuits in order is to find the average performance for an arbitrary input. In the case of sequential search, the worst and best times can be produced intentionally if the search algorithm is known. On the other hand, if we search randomly, we can expect the average time to be the same for all inputs. If the number of quantum gates used is $N$ and the depth of the quantum circuit is $d$, then the search space is $N^d$. The search space increases exponentially with the depth of the quantum circuit, and we performed a random search as a baseline to show how long it would take if we computed it.

\subsubsection{Neural Network Guided Search}

\begin{figure}[h]
    \centering
    \includegraphics[width=1.0\linewidth]{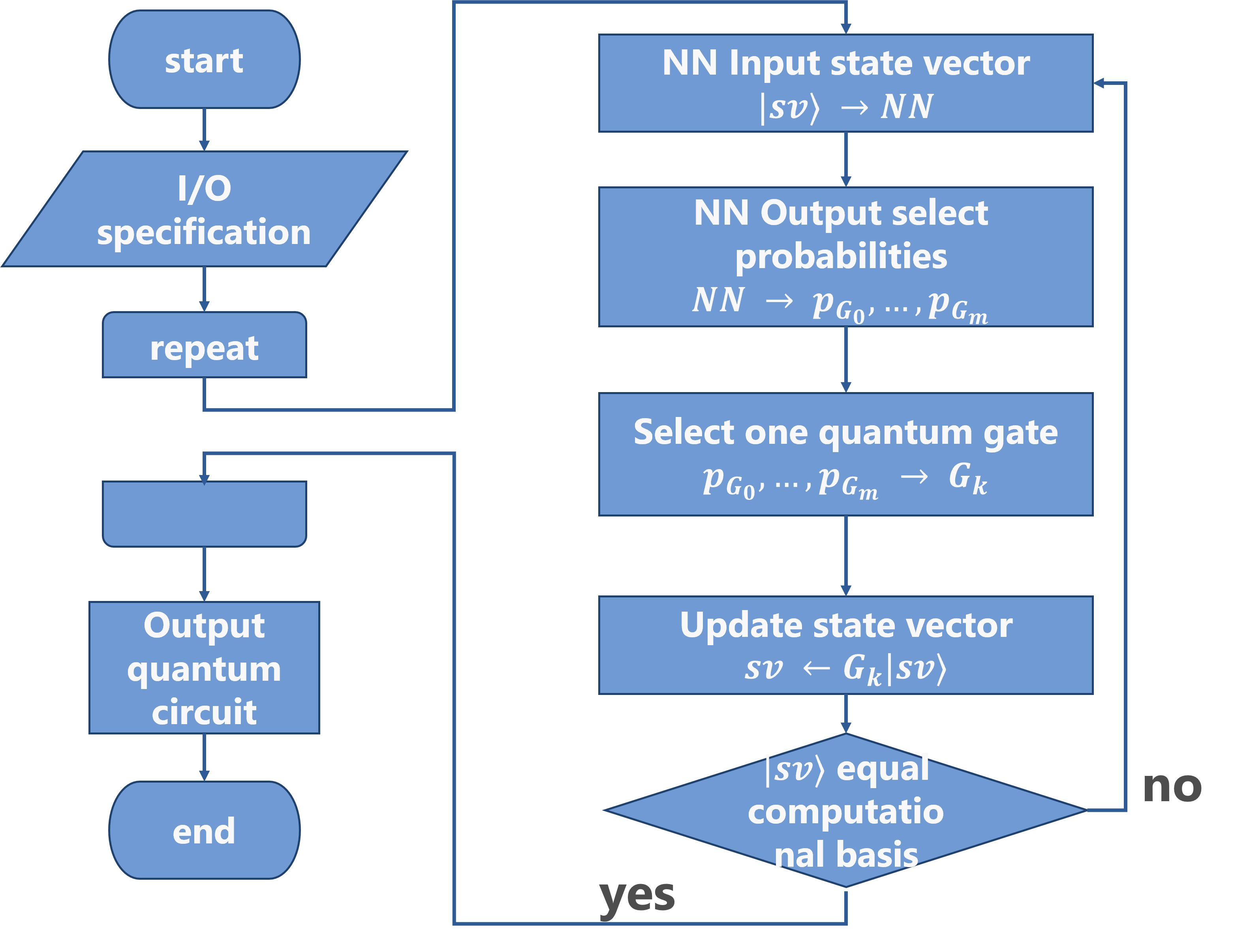}
    \caption{Synthesis algorithm: neural network guided synthesis.}
    \label{fig:algorithm}
\end{figure}

In the neural network-guided search which we propose in this paper, a neural network is used to prioritize the generation probability of quantum gates. The neural network inputs the state vector after applying the quantum gates generated up to that point in reverse order to the final state and outputs the selected probability for each quantum gate. Based on the probabilities, the next quantum gate is selected. Moreover, the conjugate transposition of the quantum gate is applied to the current state to obtain the next quantum state. This process is repeated a certain number of times until the state is on a computational basis. If the desired quantum circuit cannot be generated, it returns to the beginning. The description of the algorithm is shown in Figure~\ref{fig:algorithm}.

\subsection{Neural Network}

\subsubsection{The Structure of Network}

\begin{figure*}[tbh]
    \centering
    \includegraphics[width=0.7\linewidth]{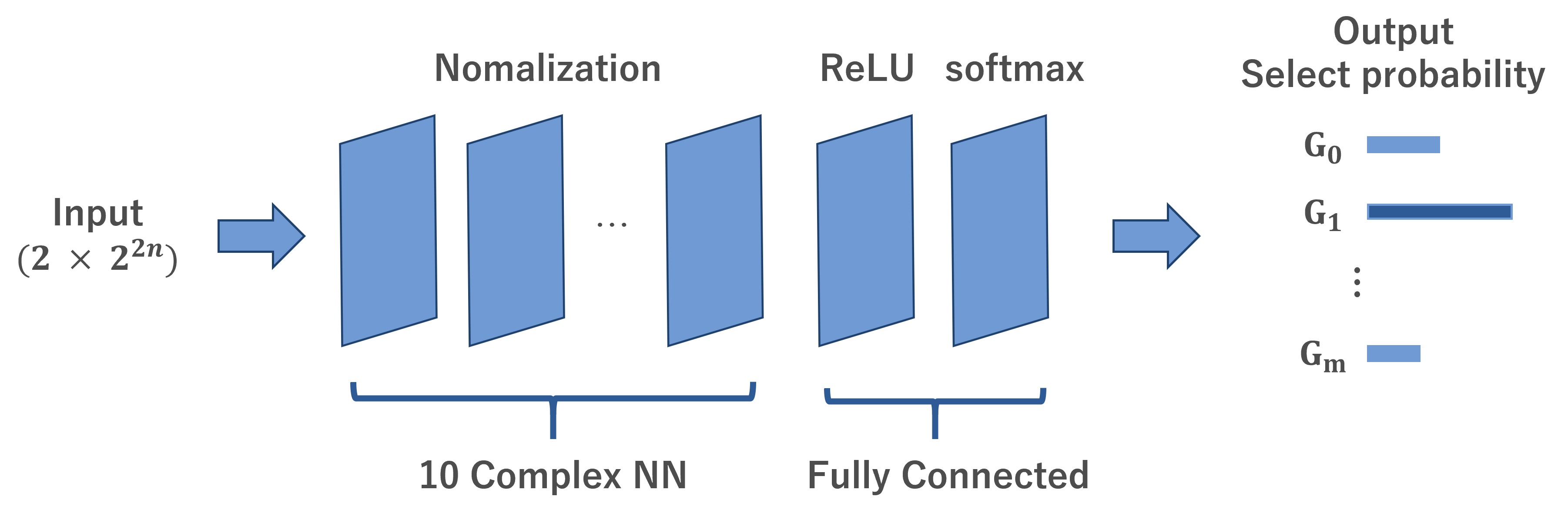}
    \caption{Structure of network. The network consists of a 10-layer complex neural network and two fully connected layers.}
    \label{fig:network}
\end{figure*}

The structure of the neural network is shown in Figure~\ref{fig:network}. The network consists of a 10-layer CvNN (complex-valued neural network) and two fully connected layers. A CvNN has real and imaginary weights and performs complex number calculations during inference and loss calculation~\cite{hirose2012complex}. In contrast to CvNN, general neural networks are sometimes called "real neural networks." The word "real" means a real number.

The advantage of CvNN is that it can increase the expressiveness of periodic motions. Specifically, it can handle shifts and rotations, at which real neural networks are not good. We adopt a CvNN since the quantum operation is unitary, corresponding to a rotation on the hypersphere in complex space. The fully connected layer at the back is a layer that maps complex numbers to real numbers and outputs the probability of selecting a quantum gate.

\subsubsection{Generating Training Data}

A neural network requires a large amount of data to function effectively. In this case, a large number of quantum circuits are required, but designing quantum circuits is a costly task for humans. For this reason, we generate the training data automatically. 

As mentioned above, a neural network takes a quantum state as input and outputs a quantum gate that looks promising at the time. Such input-output pairs are extracted from randomly generated quantum circuits. At this time, the generated quantum circuits should not overlap with other quantum circuits. Moreover, quantum circuits that loop quantum states, such as applying H-gates to the same qubit twice in a row, are eliminated.

\begin{figure}[tbh]
\large{
    \centering
    \begin{tabular}{cc}
        \begin{minipage}{.7\linewidth}
    
            \centering
            \[
            \Qcircuit @C=1.2em @R=1em{
            \lstick{A} & \gate{H} & \ctrl{1} & \gate{H} & \rstick{O_0} \qw \\
            \lstick{B} & \gate{H} & \targ    & \gate{H} & \rstick{O_1} \qw
            }\]
            \subcaption{Circuit.}
            \label{fig:multi correct circuit}
        \end{minipage}%
\\
\\
        \begin{minipage}{.7\linewidth}
            \centering
            \includegraphics[width=1.1\linewidth]{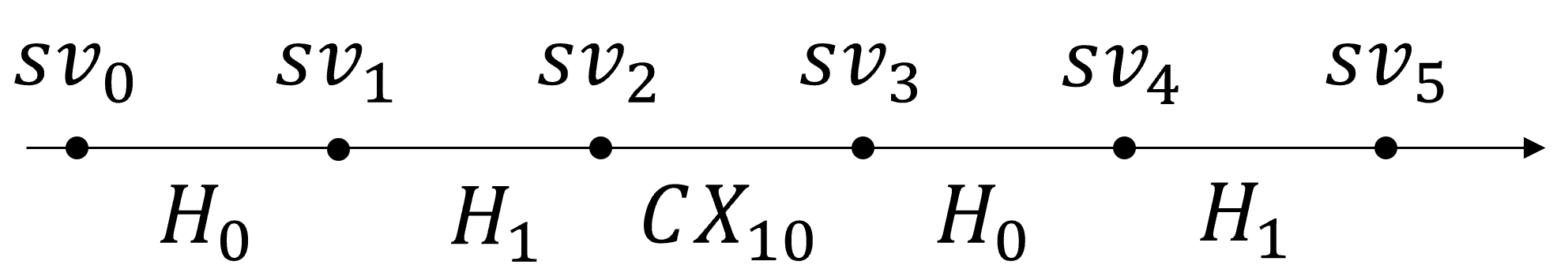}
            \subcaption{Sequence.}
            \label{fig:multi correct seq}
        \end{minipage}
    \end{tabular}
    \caption{Example of a quantum circuit with multiple correct quantum gates (left), and its sequence representation (right). The two H-gates can be generated from either side.}
    \label{fig:multi correct}
}
\end{figure}

Figure~\ref{fig:multi correct circuit} gives an example of a quantum circuit with multiple quantum gates. We use it to describe the input and output of the neural network. The input and output are denoted by $sv$ and $p$, respectively, where $sv$ is a vector of quantum states at a certain time, and $p$ is a vector of gates to be generated next. In this example, 
$p$ is represented as follows:
$$p = (p_{H_0}, p_{H_1}, p_{CX_{01}}, p_{CX_{10}})$$

$p_{G_i}$ is 1 for the next gate to be generated, and 0 otherwise, and $G_i$ denotes which gate G is applied to which qubit $i$. Since circuit~\ref{fig:multi correct circuit} is represented as a sequence, as shown in Figure~\ref{fig:multi correct seq}, the input and output of the neural network are as follows:
$$
(sv_5, (0, 1, 0, 0)), \cdots, (sv_1, (1, 0, 0, 0))
$$

Note that there may be quantum gates that can be generated from either side, as shown in Figure~\ref{fig:multi correct}. In such a case, all combinations are considered. The set of candidate quantum gates for the correct answer is $T$, and $ T'$ denotes an arbitrary subset of $T$. Let $y$ be the difference set of $T$ and $T'$, $y$ are the correct quantum gates. The input is the state after applying the quantum gate of $y$ to the state at that time, and the output is $y$. Therefore, the input and output of the neural network are as follows:
$$
(sv_5, (1, 1, 0, 0)), (sv_4, (1, 0, 0, 0)), (sv_3, (0, 1, 0, 0)), \cdots
$$

Next, we describe how to find the set of quantum gates that are candidates for the correct answer. We use the property of quantum circuits as DAG (Directed Acyclic Graph). A quantum circuit is a DAG if the quantum gates are represented as nodes and the input-output relationship as an edge. A correct quantum gate is not an input to another quantum gate. In other words, a node without outgoing edges is a correct quantum gate.
In this way, several input-output pairs can be extracted from a single quantum circuit.

\subsubsection{Learning Neural Network}

Using the above method, we randomly generated 100,000 quantum circuits (quantum cost: 10) and extracted the input-output pairs. The quantum cost is the number of two-qubit gates used in the quantum circuit. The number of extracted input-output pairs was approximately 1.13 million. Using this training data, we trained a neural network with a batch size of 64 and several epochs of 40.

\section{Experimental Evaluation}\label{sec:experimental}

To evaluate our approach, we implemented an $AutoQC$ prototype and acquired seven benchmark quantum circuits to evaluate this prototype. Each quantum circuit is derived from the widely used quantum circuits in the community. For each quantum circuit, we generated input-output pairs automatically, fed these pairs into the $AutoQC$ prototype, and obtained the output results, including the topology of the synthesized quantum circuit.

Our experiments were performed on a system with AMD Ryzen 5 5600X 6-Core Processor 3.70 GHz, 32GB RAM, NVIDIA GeForce RTX 3060 ti. 

\subsection{A Benchmark of Quantum Circuits for Synthesis}
We use a benchmark consisting of the following quantum circuits for performing our synthesis experiment. These quantum circuits are important building blocks for the implementation of reversible logic circuits, more complex quantum multiplication circuits, and quantum ALUs~\cite{ALU1,ALU2} because they can simultaneously compute various logic functions, including full addition.

In the following, we briefly describe each circuit, and for more detailed descriptions of these quantum circuits, please refer to the related papers.

\vspace*{3mm}
\begin{itemize}[leftmargin=2em,itemsep=0.5em]
\item \textbf{HNG}~\cite{HNG}: HNG is a 4-input, 4-output reversible logic gate that can be used to work singly as a reversible full adder circuit when its fourth input is set to zero.

\item \textbf{PFAG}~\cite{PFAG}: PFAG gate is almost similar to HNG except that PFAG provides half adder sum output besides providing full adder sum output. 

\item \textbf{IG}~\cite{IG}: IG gate is used to construct a Full Adder-Subtractor circuit.
\item \textbf{MIG}~\cite{MIG}: MIG gate is the modified version of the IG gate. MIG gate is used to design carry look-ahead adder.
\item \textbf{OTG}~\cite{OTG}: OTG gate is suitable for online testability in reversible logic circuits.
\item \textbf{MKG}~\cite{MKG}: MKG gate is used to construct the reversible multiplier circuit.
\item \textbf{TSG}~\cite{TSG1,TSG2}: TSG gate is used to design ripple carry adder, BCD adder, and the carry look-ahead adder.
\end{itemize}
\vspace*{3mm}


\subsection{Experimental Results}
First, we conducted a random search experiment as a baseline, which gives a rough idea of how complex the quantum circuit synthesis problem is. The HNG in Figure~\ref{fig:HNG} was searched for one hour by random search, and the number of quantum circuits seen during that time was used to estimate the running time of the entire search. In our environment, it took about 169 hours to generate a quantum circuit with four qubits and a quantum cost of 6 (the order of the search space is about $10^9$) with 50\% probability. 

We next conducted an experiment using our approach to generate quantum circuits from our benchmark based on their input-output pairs.
%
This approach can be easily applied to these benchmarks and other quantum circuits with publicly available input and output specifications.

The results of the proposed approach can be seen in Table~\ref{tab:result20}, which shows the results of 100 trials. Moreover, the mean and standard deviation of running time and the number of circuits explored to find the desired quantum circuit, and the minimum quantum cost is shown. The quantum cost is the number of two-qubit gates used in the quantum circuit. TSG is shown as a failure in the table because the search was terminated after 1,000 trials based on the number of search spaces for other quantum circuits.

\begin{table*}[tb]
\centering
\caption{Results of 100 experiments performed on a network trained on quantum circuits with quantum cost 10 (AutoQC10) and quantum cost 20 (AutoQC20). We could not calculate the mean and standard deviation of TSG because we could only synthesize TSG once.}
\label{tab:result10}
\scalebox{1.0}{
\begin{tabular}{@{}cccccccccc@{}}
\toprule
\multicolumn{1}{l}{} & \multicolumn{4}{c}{Quantum Cost}                    & \multicolumn{3}{c}{Time[sec]}          & \multicolumn{2}{c}{Loop} \\ \cmidrule(r){2-10}
Circuit              & AutoQC10 & AutoQC20 & \cite{velasquez2021automated} & \cite{various_gate} & AutoQC10 & AutoQC20 & \cite{velasquez2021automated} & AutoQC10 & AutoQC20 \\ \midrule
HNG     & 6              & 6          & 6       & 6     & 2.67 $\pm$ 2.06  & 3.98 $\pm$ 3.21  & 224.50     & 131.68 $\pm$ 105.60 & 184.95 $\pm$ 155.56 \\
PFAG    & 6              & 6          & 6       & -     & 5.50 $\pm$ 3.97  & 6.30 $\pm$ 4.94  & 193.92     & 284.43 $\pm$ 213.60 & 320.72 $\pm$ 260.54 \\
MIG     & \textbf{7}     & \textbf{7} & -       & 8     & 8.60 $\pm$ 6.41  & 3.73 $\pm$ 3.06  & -          & 325.12 $\pm$ 250.44 & 128.22 $\pm$ 108.73 \\
IG      & \textbf{7}     & \textbf{7} & -       & 9     & 8.96 $\pm$ 7.82  & 2.40 $\pm$ 2.88  &-           & 293.38 $\pm$ 266.00 & 72.40 $\pm$ 91.10 \\
OTG     & \textbf{8}     & \textbf{8} & -       & 10    & 6.29 $\pm$ 5.70  & 13.25 $\pm$ 9.26 &-           & 176.16 $\pm$ 165.22 & 368.91 $\pm$ 262.92 \\
MKG     & \textbf{9}     & \textbf{9} & 15      & 17    & 7.31 $\pm$ 7.29  & 4.15 $\pm$ 4.03  & 159,473.84 & 120.38 $\pm$ 123.64 & 64.45 $\pm$ 64.15 \\
TSG     & fail           & 19         & 13      & 20    & -                & 15.06            & 144,326.49 & 1000 & 183             \\ \bottomrule
\end{tabular}
}
\end{table*}

The search is clearly more efficient than a random one, which suggests that the neural network can select the appropriate quantum gate. In addition, although there is no comparison, the number of searches is small enough.

A comparison with the previous study is also made in Table~\ref{tab:result10}. In the previous study, SWAP gates can be used unlike our approach, but since SWAP gates can be created with three CX gates, SWAP gates are counted as three CX gates. Since the experimental environment is different from the previous study, a simple comparison of execution time is not possible. However, except for TSG, which failed to generate, our approach was faster and achieved the lowest quantum cost.

\begin{figure}[tb]
    \centering
    \includegraphics[width=1.0\linewidth]{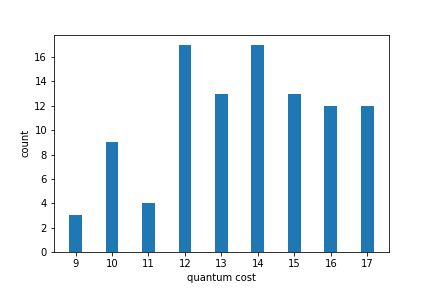}
    \caption{Distribution of quantum cost of synthesized MKG using a network trained on cost 10 circuits.}
    \label{fig:MKG_dist}
\end{figure}

The distribution of the quantum cost of synthesized MKG is shown in Figure~\ref{fig:MKG_dist}. The quantum cost distributions of other quantum circuits and the generated quantum circuits are summarized in Appendix A. It is noteworthy that, contrary to our expectations, we could synthesize a quantum circuit with a cost greater than the trained cost. 

In addition, to synthesize TSG that could not be synthesized at the cost 10 model, we generated training data and trained the neural network in a circuit at cost 20. The number of input-output pairs is 2.28 million, about twice as many as the cost of 10 cases. The results of the cost 20 model are also shown in Table~\ref{tab:result20}. 

As a result, we succeeded in synthesizing only one out of 100 trials. Moreover, the cost 20 model is generally better in terms of loop count. Therefore, the accuracy may be further improved by increasing the cost and the training data.

\begin{table*}[tbh]
\centering
\caption{Results of 100 experiments performed on a network trained on quantum circuits with cost 20. We could not calculate the mean and standard deviation because we could only synthesize TSG once.}    
\label{tab:result20}
\scalebox{1.0}{
\begin{tabular}{@{}ccccccc@{}}
\toprule
\multicolumn{1}{l}{} & \multicolumn{2}{c}{Quantum Cost} & \multicolumn{2}{c}{Time[sec]} & \multicolumn{2}{c}{Loop} \\ \cmidrule(r){2-7}
Circuit              & cost10 & cost20 & cost10 & cost20  & cost10 & cost20 \\ \midrule
HNG     & 6 & 6           & 2.67 $\pm$ 2.06  & 3.98 $\pm$ 3.21 & 131.68 $\pm$ 105.60 & 184.95 $\pm$ 155.56 \\
PFAG    & 6 & 6           & 5.50 $\pm$ 3.97  & 6.30 $\pm$ 4.94  & 284.43 $\pm$ 213.60 & 320.72 $\pm$ 260.54 \\
MIG     & 7 & 7           & 8.60 $\pm$ 6.41  & 3.73 $\pm$ 3.06  & 325.12 $\pm$ 250.44 & 128.22 $\pm$ 108.73 \\
IG      & 7 & 7           & 8.96 $\pm$ 7.82  & 2.40 $\pm$ 2.88  & 293.38 $\pm$ 266.00 & 72.40 $\pm$ 91.10 \\
OTG     & 8 & 8           & 6.29 $\pm$ 5.70  & 13.25 $\pm$ 9.26  & 176.16 $\pm$ 165.22 & 368.91 $\pm$ 262.92 \\
MKG     & 9 & 9           & 7.31 $\pm$ 7.29  & 4.15 $\pm$ 4.03  & 120.38 $\pm$ 123.64 & 64.45 $\pm$ 64.15 \\
TSG     & - & 19          & -  & 15.06           & - & 183               \\ \bottomrule
\end{tabular}
}
\end{table*}

The problem with our approach is that even if we succeed in synthesizing a quantum circuit in a single trial, we do not know if it is an optimal quantum circuit. Therefore, it is necessary to select a suitable circuit after several trials, as in this experiment.

\section{Related work}
\label{sec:related work}

Cruz-Benito {\it et al.}~\cite{cruz2018deep} use deep learning to learn how people are using the OpenQASM language and provide help and guidance to coders by recommending different code sequences, logical steps, or small pieces of code. They use a seq2seq neural network to learn the code of a quantum programming language and predict the next sentence based on the previous sentences written.

Several automatic synthesis approaches for quantum circuits have been proposed using genetic algorithms. Ruican {\it et al.}~\cite{ruican2007automatic} proposed an automatic quantum circuit synthesis approach using genetic algorithms. The basic idea of the approach is to divide the potential circuits into vertical levels called "sections" and horizontal levels called "planes" to achieve a consistent representation of genetic algorithms for chromosome definition. They also suggest using a database to specify the gates used in the synthesis process. Lukac and Perkowski~\cite{lukac2002evolving} proposed an approach to evolve quantum circuits using genetic algorithms. The genetic algorithm automatically searches for an appropriate circuit design that produces the desired output state. The fitness function compares the current output to the desired output and stops the search when a close match is found.


Velasquez {\it et al.}~\cite{velasquez2021automated} proposed an approach to automatically synthesizing a quantum circuit using a decision procedure that performs symbolic reasoning for combinatorial search. This automatic synthesis approach automatically constructs a finite symbolic abstract model of the quantum gate and generates constraints from this symbolic model and the input-output pairs of the quantum circuit. An SMT solver then solves the constraints to discover the quantum circuits. They demonstrate the potential of this approach by automatically synthesizing four quantum circuits and rediscovering the Bernstein-Vazirani quantum algorithm~\cite{bernstein1997quantum}.

Compared to the above approaches, our approach uses a neural network-based approach to synthesizing quantum circuits automatically, which is more efficient and at a lower cost and shows that neural networks can effectively support the automated development of quantum circuits.

\section{Concluding Remarks}\label{sec:conclusion}

This paper has proposed an approach to automatically synthesizing quantum circuits using a neural network from input-output pairs. Our approach can generate quantum circuits at a lower cost, which shows that neural networks can effectively support the development of quantum circuits. 

In the future, we would like to investigate how well the proposed approach can scale up in future work. We have shown that the algorithm scales to some extent, but it is necessary to investigate increasing the number of qubits, the depth of the quantum circuit, and training data. In addition, since the input size of the algorithm increases in proportion to the exponent of the number of qubits, an efficient way to represent the quantum state is also required. 

\bibliographystyle{IEEEtran}
\bibliography{sotsuron}

\newpage
\appendix
\renewcommand{\appendixname}{Appendix~\Alph{section}}
\noindent

\vspace*{2mm}
Figures 8 $\sim$ 15 show some examples of quantum circuits synthesized by our approach and the distribution of quantum costs.

\begin{figure}[thb]
\centering
    \begin{tabular}{cc}
        \begin{minipage}[t]{.4\linewidth}
            \centering
            \includegraphics[width=1.2\linewidth]{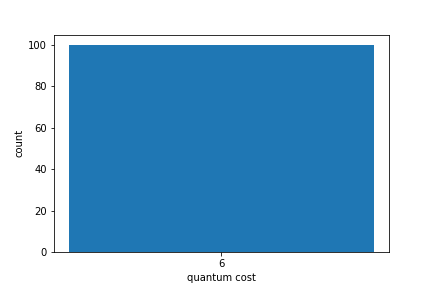}
            \subcaption{HNG}
        \end{minipage}%
        \begin{minipage}[t]{.4\linewidth}
            \centering
            \includegraphics[width=1.2\linewidth]{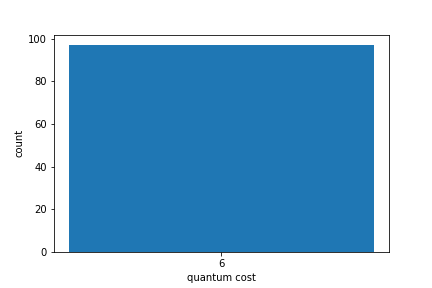}
            \subcaption{PFAG}
        \end{minipage}\\
        \begin{minipage}[t]{.4\linewidth}
            \centering
            \includegraphics[width=1.2\linewidth]{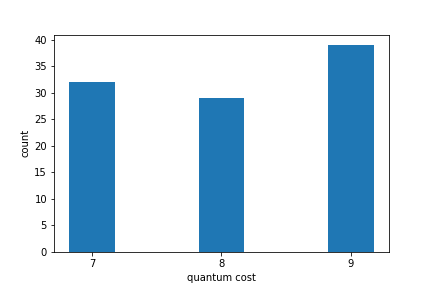}
            \subcaption{IG}
        \end{minipage}
        \begin{minipage}[t]{.4\linewidth}
            \centering
            \includegraphics[width=1.2\linewidth]{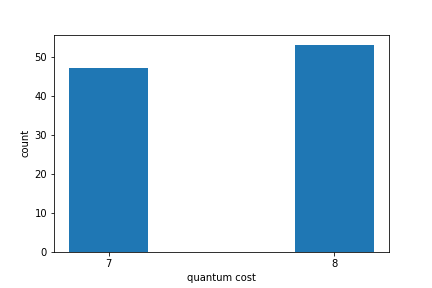}
            \subcaption{MIG}
        \end{minipage}\\%
        \begin{minipage}[t]{.4\linewidth}
            \centering
            \includegraphics[width=1.2\linewidth]{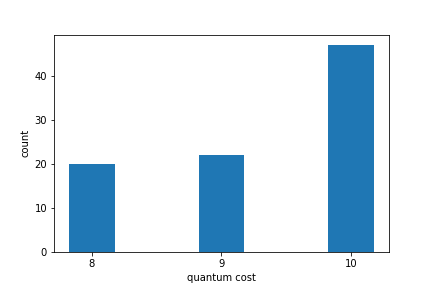}
            \subcaption{OTG}
        \end{minipage}%
        \begin{minipage}[t]{.4\linewidth}
            \centering
            \includegraphics[width=1.2\linewidth]{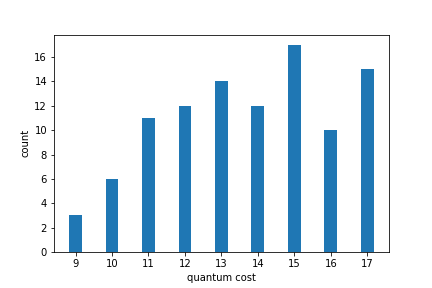}
            \subcaption{MKG}
        \end{minipage}\\
    \end{tabular}
    \caption{Distribution of quantum cost of the synthesized quantum circuits using a network trained on depth 20 circuits.}\label{fig:distribution}
\end{figure}

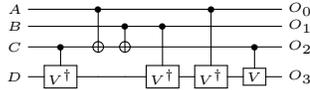
\begin{figure}[h]
\tiny{
    \centering
    \[
    \Qcircuit @C=1.2em @R=1em{
        \lstick{A} & \qw              & \ctrl{2} & \qw      & \qw                 & \ctrl{3}              & \qw      & \rstick{O_0} \qw \\
        \lstick{B} & \qw              & \qw      & \ctrl{1} & \ctrl{2}            & \qw                   & \qw      & \rstick{O_1} \qw \\
        \lstick{C} & \ctrl{1}         & \targ    & \targ    & \qw                 & \qw                   & \ctrl{1} & \rstick{O_2} \qw \\
        \lstick{D} & \gate{V^\dagger} & \qw      & \qw      & \gate{V^\dagger}    & \gate{V^\dagger}      & \gate{V} & \rstick{O_3} \qw \\
    }\]
    \caption{HNG synthesized by depth 20 model. A, B, C and D are the inputs, and $O_0=A$, $O_1=B$, $O_2=A\oplus B\oplus C$ and $O_3=(A\oplus B)C\oplus AB\oplus D$ are the outputs.}
    \label{fig:HNG2}
}
\end{figure}

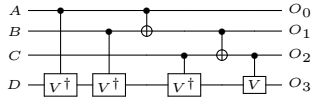
\begin{figure}[h]
\tiny{
    \centering
    \[
    \Qcircuit @C=1.2em @R=1em{
        \lstick{A} & \ctrl{3}         & \qw              & \ctrl{1} & \qw                 & \qw      & \qw      & \rstick{O_0} \qw \\
        \lstick{B} & \qw              & \ctrl{2}         & \targ    & \qw                 & \ctrl{1} & \qw      & \rstick{O_1} \qw \\
        \lstick{C} & \qw              & \qw              & \qw      & \ctrl{1}            & \targ    & \ctrl{1} & \rstick{O_2} \qw \\
        \lstick{D} & \gate{V^\dagger} & \gate{V^\dagger} & \qw      & \gate{V^\dagger}    & \qw      & \gate{V} & \rstick{O_3} \qw \\
    }\]
    \caption{PFAG synthesized by depth 20 model. $O_0=A$, $O_1=A\oplus B$, $O_2=A\oplus B\oplus C$ and $O_3=(A\oplus B)C \oplus AB \oplus D$ are the outputs.}
    \label{fig:PFAG}
}
\end{figure}

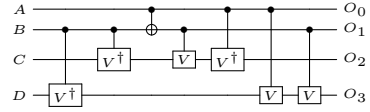
\begin{figure}[h]
\tiny{
    \centering
    \[
    \Qcircuit @C=1.2em @R=1em{
        \lstick{A} & \qw              & \qw              & \ctrl{1}  & \qw      & \ctrl{2}         & \ctrl{3} & \qw      & \rstick{O_0} \qw \\
        \lstick{B} & \ctrl{2}         & \ctrl{1}         & \targ     & \ctrl{1} & \qw              & \qw      & \ctrl{2} & \rstick{O_1} \qw \\
        \lstick{C} & \qw              & \gate{V^\dagger} & \qw       & \gate{V} & \gate{V^\dagger} & \qw      & \qw      & \rstick{O_2} \qw \\
        \lstick{D} & \gate{V^\dagger} & \qw              & \qw       & \qw      & \qw              & \gate{V} & \gate{V} & \rstick{O_3} \qw \\
    }\]
    \caption{IG synthesized by depth 20 model. $O_0=A$, $O_1=A \oplus B$, $O_2=AB \oplus C$ and $O_3=DB \oplus \bar{B} (A \oplus D)$ are the outputs.}
    \label{fig:IG}
}
\end{figure}

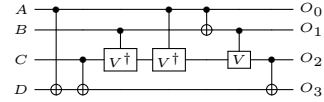
\begin{figure}[h]
\tiny{
    \centering
    \[
    \Qcircuit @C=1.2em @R=1em{
        \lstick{A} & \ctrl{3} & \qw      & \qw              & \ctrl{2}         & \ctrl{1} & \qw      & \qw      & \rstick{O_0} \qw \\
        \lstick{B} & \qw       & \qw      & \ctrl{1}         & \qw              & \targ    & \ctrl{1} & \qw      & \rstick{O_1} \qw \\
        \lstick{C} & \qw       & \ctrl{1} & \gate{V^\dagger} & \gate{V^\dagger} & \qw      & \gate{V} & \ctrl{1} & \rstick{O_2} \qw \\
        \lstick{D} & \targ     & \targ    & \qw              & \qw              & \qw      & \qw      & \targ    & \rstick{O_3} \qw \\
    }\]
    \caption{MIG synthesized by depth 20 model. $O_0=A$, $O_1=A \oplus B$, $O_2=AB \oplus C$ and $O_3=A\bar{B} \oplus D$ are the outputs.}
    \label{fig:MIG}
}
\end{figure}

\begin{figure}[t]
\tiny{
    \centering
    \[
    \Qcircuit @C=1.2em @R=1em{
        \lstick{A} & \qw      & \qw              & \qw      & \ctrl{1}  & \qw       & \qw      & \ctrl{3} & \qw              & \rstick{O_0} \qw \\
        \lstick{B} & \ctrl{1} & \qw              & \qw      & \targ     & \qw       & \ctrl{1} & \qw      & \qw              & \rstick{O_1} \qw \\
        \lstick{C} & \gate{V} & \gate{V^\dagger} & \ctrl{1} & \qw       & \targ     & \targ    & \qw      & \ctrl{1}         & \rstick{O_2} \qw \\
        \lstick{D} & \qw      & \ctrl{-1}        & \targ    & \qw       & \ctrl{-1} & \qw      & \gate{V} & \gate{V^\dagger} & \rstick{O_3} \qw \\
    }\]
    \caption{OTG synthesized by depth 20 model. $O_0=A$, $O_1=A \oplus B$, $O_2=A \oplus B \oplus D$ and $O_3=(A \oplus B)D \oplus AB \oplus C)$ are the outputs.}
    \label{fig:OTG}
}
\end{figure}
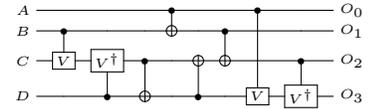

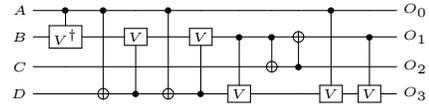
\begin{figure}[t]
\tiny{
    \centering
    \[
    \Qcircuit @C=1.2em @R=1em{
        \lstick{A} & \ctrl{1}        & \ctrl{3} & \qw       & \ctrl{3} & \qw       & \qw      & \qw      & \qw       & \ctrl{3} & \qw      & \rstick{O_0} \qw \\
        \lstick{B} & \gate{V^\dagger} & \qw      & \gate{V}  & \qw      & \gate{V}  & \ctrl{2} & \ctrl{1} & \targ     & \qw      & \ctrl{2} & \rstick{O_1} \qw \\
        \lstick{C} & \qw             & \qw      & \qw       & \qw      & \qw       & \qw      & \targ    & \ctrl{-1} & \qw      & \qw      & \rstick{O_2} \qw \\
        \lstick{D} & \qw             & \targ    & \ctrl{-2} & \targ    & \ctrl{-2} & \gate{V} & \qw      & \qw       & \gate{V} & \gate{V} & \rstick{O_3} \qw \\
    }\]
    \caption{MKG synthesized by depth 20 model. $O_0=A$, $O_1=C$, $O_2=(\bar{A}\bar{D} \oplus \bar{B}) \oplus C$ and $O_3=(\bar{A}\bar{D} \oplus \bar{B})C \oplus (AB \oplus D)$ are the outputs.}
    \label{fig:MKG}
}
\end{figure}

\begin{figure}[h]
\tiny{
    \centering
    \[
    \Qcircuit @C=.6em @R=1em{
        \lstick{A} & \ctrl{1} & \ctrl{2} & \ctrl{2} & \qw & \ctrl{2} & \qw & \qw & \qw & \qw & \qw & \ctrl{1} & \qw & \qw & \qw & \qw & \ctrl{2} & \ctrl{2} & \qw & \qw & \rstick{O_0} \qw \\
        \lstick{B} & \gate{V} & \qw & \qw & \gate{V^\dagger} & \qw & \gate{V} & \targ & \ctrl{1} & \qw & \ctrl{1} & \targ & \ctrl{1} & \ctrl{2} & \qw & \qw & \qw & \qw & \qw & \qw & \rstick{O_1} \qw \\
        \lstick{C} & \qw & \targ & \gate{V^\dagger} & \ctrl{-1} & \targ & \ctrl{-1} & \ctrl{-1} & \gate{V} & \gate{V} & \targ & \qw & \targ & \qw & \gate{V^\dagger} & \ctrl{1} & \gate{V} & \gate{V^\dagger} & \ctrl{1} & \ctrl{1} & \rstick{O_2} \qw \\
        \lstick{D} & \qw & \qw & \qw & \qw & \qw & \qw & \qw & \qw & \ctrl{-1} & \qw & \qw & \qw & \targ & \ctrl{-1} & \gate{V^\dagger} & \qw & \qw & \gate{V} & \targ
 & \rstick{O_3} \qw \\
    }\]
    \caption{TSG synthesized by depth 20 model. $O_0=A$, $O_1=\bar{A}\bar{C}\oplus \bar{B}$, $O_2=(\bar{A}\bar{C}\oplus \bar{B}) \oplus D$ and $O_3=(\bar{A}\bar{C}\oplus \bar{B})D \oplus AB \oplus C$ are the outputs.}
    \label{fig:TSG}
}
\end{figure}
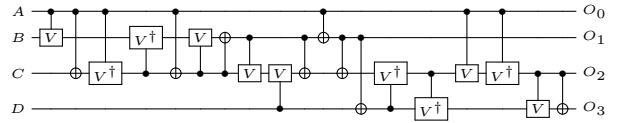





\end{document}